\documentclass[a4paper, 12pt]{article}

\usepackage[margin=1in]{geometry}
\usepackage[mathlines]{lineno}                        

\usepackage{amssymb,amsmath,amsthm,amsfonts,bbm,graphicx,subfigure}
\usepackage[sort,comma,numbers]{natbib}

\def\boxit#1{\hsize=2truein
        \vbox{\hrule\hbox{\vrule\kern3pt
      \vbox{\kern3pt#1\kern3pt}\kern3pt\vrule}\hrule}}

\newcommand{\cF}{\mathcal{F}}
\newcommand{\dd}{\textup{d}}
\newcommand{\E}{\mathbb{E}}
\newcommand{\empk}{\widehat{G}_{\hspace{-0.2mm} k}}
\newcommand{\emps}{\widehat{G}_{\hspace{-0.2mm} 1}}
\newcommand{\hsp}{{\hspace{0.30mm}}} 
\newcommand{\real}{\mathbb{R}}
\newcommand{\bl}{\begin{linenomath}} 
\newcommand{\el}{\end{linenomath}}

\begin{document}

\title{Using proper divergence functions to evaluate \\ climate models}

\author{Thordis L. Thorarinsdottir\footnote{Norwegian Computing Center, PO Box 114, 
                                            Blindern, 0314 Oslo, Norway},
Tilmann Gneiting\footnote{Institut f\"ur Angewandte Mathematik, Universit\"at Heidelberg, 
                          Im Neuenheimer Feld 294, 69120 Heidelberg, Germany}, 
and Nadine Gissibl\footnote{Lehrstuhl f\"ur Mathematische Statistik, Zentrum Mathematik, Technische Universit\"at
M\"unchen, Parkring 13, 85748 Garching, Germany}}

\maketitle

\begin{abstract} 
\noindent
It has been argued persuasively that, in order to evaluate climate
models, the probability distributions of model output need to be
compared to the corresponding empirical distributions of observed
data.  Distance measures between probability distributions, also
called divergence functions, can be used for this purpose.  We contend
that divergence functions ought to be proper, in the sense that acting
on modelers' true beliefs is an optimal strategy.  Score divergences
that derive from proper scoring rules are proper, with the integrated
quadratic distance and the Kullback-Leibler divergence being
particularly attractive choices.  Other commonly used divergences fail
to be proper.  In an illustration, we evaluate and rank simulations
from fifteen climate models for temperature extremes in a comparison
to re-analysis data.
\end{abstract}

\section{Introduction}  \label{sec:introduction}

Traditionally, climate models have been assessed by
comparing summary statistics or point estimates that derive from the
simulated model output to the corresponding observed quantities
\cite{Gleckler&2008, Mearns&2012}.  However, there is a growing
consensus that in order to evaluate climate models the probability
distribution of model output needs to be compared to the corresponding
distribution of observed data.  Guttorp \cite[p.~820]{Guttorp2011}
argues powerfully that

\begin{quote}
\footnotesize
Climate models are difficult to compare to data. Often
climatologists compute some summary statistic, such as global annual
mean temperature, and compare climate models using observed (or rather
estimated) forcings to the observed (or rather estimated)
temperatures. However, it seems more appropriate to compare the
distribution (over time and space) of climate model output to the
corresponding distribution of observed data, as opposed to point
estimates with or without confidence intervals.
\end{quote} 

\noindent 
Palmer \cite[p.$\:$850]{Palmer2012} and Arnold et
al.~\cite{Arnold&2013} join this argument, by proposing the use of the
Hellinger distance between simulated and observed probability
distributions to assess climate models.  A pioneering effort in these
directions is that of Perkins et al.$\;$\cite{Perkins&2007} who use a
transportation type metric; however, this metric applies to densities
only and thus requires density estimates.

Taking a broader perspective, predictive models are key tools in many
realms of science and society, and principled methods for assessing
their performance are in strong demand.  Any such evaluation procedure
ought to provide a quantitative assessment of the compatibility of the
simulation model, and the real world phenomena it is meant to
represent, in a manner that encourages careful assessment and
integrity.  In this paper, we focus on the setting in which the
simulation model supplies a probability distribution, $F$, typically
consisting of model output composited over time and/or space, and we
wish to evaluate it against the empirical measure, $\empk$, associated 
with the corresponding observations.  The evaluation then is usually 
performed by computing a distance measure or divergence, $d$, between 
the probability distributions $F$ and $\empk$, where the divergence
function $d$ is such that $d(F,F) = 0$ and $d(F,G) \geq 0$ for all
probability distributions $F$ and $G$.  Competing climate models give
rise to distinct simulated probability distributions, and those that
yield the smallest divergences when compared to the empirical measure
are considered to perform best. 

Myriads of distance measures for probability distributions have been
studied in the literature, as reviewed comprehensively by Deza and
Deza \cite[Chapter 14]{Deza&2013} and Gibbs and Su \cite{GibbsSu2002},
and it is far from obvious which ought to be used for evaluation
purposes.  Our paper aims to provide guidance in the choice of the
divergence function in a general setting, including both categorical
and continuous, real- or vector-valued observations.  The task is not
unique to climatology, and essentially the same problem arises in the
emerging transdisciplinary field of uncertainty quantification.  In
this latter context, Ferson et al.$\;$\cite[\S 5.2]{Ferson&2008} list
desirable properties of a divergence function $d$ in the case of a
real-valued observation, in that\footnote{Ferson et al.$\;$\cite[\S
5.2]{Ferson&2008} furthermore request the unboundedness of the
divergence function.  We suppress this request, as the compatibility
property (i) implies unboundedness.}
\begin{enumerate} 
\item[(i)]
the divergence $d(\delta_x,\delta_y)$ 
between point measures $\delta_x$ and $\delta_y$ ought to reduce to the 
absolute error $|x-y|$;  
\item[(ii)]
the divergence ought to be sensitive to all distributional facets, 
rather than just means and variances; 
\item[(iii)]
the divergence ought to be expressible in physical units\footnote{For example, 
if we are concerned with temperature simulations in degrees Celsius, 
then the divergence should have a value in degrees Celsius.};  
\item[(iv)]
the divergence function ought to be ``mathematically well behaved and 
well understood''.
\end{enumerate}

\noindent 
Our goal here is to formalize property (iv), by proposing and studying
a propriety condition for divergence functions that resembles the
classical propriety condition of Winkler and Murphy
\cite{WinklerMurphy1968} for scoring rules.  Specifically, suppose
that $F$ is a predictive probability distribution for a single future
quantity or event, $y$.  In this setting, predictive performance is
assessed by assigning a numerical score, $s(F,y)$, based on the
predictive distribution, $F$, and the verifying value, $y$.  We
suppose that scoring rules are negatively oriented,
that is, the smaller, the better, and we write $s(F,G)$ for the
expectation of $s(F,y)$ when $y$ is a random variable with
distribution $G$.

The scoring rule $s$ is said to be proper if $s(G,G) \leq s(F,G)$ for
all $F, G \in \cF$, where $\cF$ is a suitable class of probability
measures.  In other words, a proper scoring rule is such that the
expected score or penalty is minimized when the predictive
distribution, $F$, agrees with the true distribution, $G$, of the
quantity or event to be forecast.  As the use of proper scoring rules
encourages honest and careful assessments,\footnote{If a
scoring rule is not proper, it may, for instance, be the case that an
artificial deflation of the predictive variance entails purported
higher skill \cite{GneitingRaftery2007}.} and is deeply entrenched in
first decision theoretic principles, it is considered essential in
scientific and managerial practice \cite{GneitingRaftery2007,
BrockerSmith2007}.

\section{Notions of Propriety for Divergences}  \label{sec:proper}

Divergence functions address situations where the prediction takes the
form of a probability distribution $F$ within a convex class, $\cF$,
of probability measures on a general sample space $\Omega$.  The
predictive distribution then needs to be compared against observations
$y_1, \ldots, y_k$ and the corresponding empirical measure,
\bl
\[
\empk = \frac{1}{k} \sum_{i=1}^k \delta_{y_i}, 
\]
\el
where $\delta_{y_i}$ is the point measure in the observation $y_i \in
\Omega$.  A {\em divergence}\/ then is a function 
\bl
\[
d: \cF \times \cF \to \real
\]
\el
such that $d(F,F) = 0$ and $d(F,G) \geq 0$ for all probability
distributions $F, G \in \cF$.  Divergences are sometimes called
divergence functions or discrepancy functions, or validation metrics
\cite{Liu&2011}, even though they are typically not metrics in the
mathematical sense.  We generally assume that the convex class $\cF$
contains all empirical measures, and we write
\bl
\[
\E_G \hsp d(F,\empk)
\]
\el
to denote the expectation of $d(F,\empk)$ when the predictive
distribution is $F \in \cF$, and the observations $y_1, \ldots, y_k
\in \Omega$ in the empirical measure are independent with distribution
$G \in \cF$.

\bigskip
\noindent{\bf Definition 1.}  
{\em A divergence function\/ $d: \cF \times \cF \to \real$ is\/
$k$-proper for a positive integer\/ $k$ if
\bl
\begin{equation}  \label{eq:k-proper}
\E_G \hsp d(G,\empk) \leq  \E_G \hsp d(F,\empk)
\end{equation}
\el
for all probability distributions\/ $F,G \in \cF$.  It is asymptotically proper if
\bl
\begin{equation}  \label{eq:asymp-proper}
\textstyle
\lim_{k \to \infty} \E_G \hsp d(G,\empk) \leq \lim_{k \to \infty} \E_G \hsp d(F,\empk),
\end{equation}
\el
for all\/ $F,G \in \cF$.}  

\bigskip
Thus, if the divergence is $k$-proper, and we believe that the
observations form a sample from the probability distribution $G$, then
an optimal strategy is to act on one's true beliefs.  In this sense,
the use of proper divergence functions encourages honest and careful
assessments.

Technically, the propriety condition (\ref{eq:k-proper}) relates
closely to the corresponding condition for scoring rules.
Recall\footnote{As noted in Section
\ref{sec:introduction}, a scoring rule is a function $s : \cF \times
\Omega \to \real$, with the first argument being a probability
distribution and the second argument being an observed value.  In
a slight abuse of notation, we write $s(F,G)$ for the
expectation of $s(F,y)$ when $y$ is a random variable with
distribution $G$.} that a scoring rule $s$ is proper if
\bl
\[
s(G,G) = \E_G \hsp s(G,y) \leq \E_G \hsp s(F,y) = s(F,G)
\]
\el
for all probability distributions $F, G \in \cF$.  Thus we see that if
the divergence $d$ is 1-proper, the scoring rule $s(F,y) =
d(F,\delta_y)$ is proper.  Conversely, if $s$ is a proper scoring rule and
$s(G,G)$ is finite for all $G \in \cF$, then
\bl
\[
d(F,G) = s(F,G) - s(G,G)
\]
\el
is a divergence function, which we refer to as the {\em score
divergence}\/ associated with the proper scoring rule $s$.

\bigskip
\noindent{\bf Theorem 2.}  
{\em If\/ $d: \cF \times \cF \to \real$ is a score divergence, then
$d$ is\/ $k$-proper for all positive integers\/ $k$.}

\begin{proof} If $d$ is a score divergence, then there exists a proper
scoring rule $s$ such that 
\bl
\begin{equation}  \label{eq:score.d}  
d(F,\empk) = s(F,\empk) - s(\empk,\empk),
\end{equation} 
\el
where
\bl
\begin{equation}  \label{eq:score}  
s(F,\empk) = \frac{1}{k} \sum_{i=1}^k s(F,y_i),  
\end{equation} 
\el
and $s(\empk,\empk)$ denotes the expected score when the predictive
distribution is the empirical measure and the observation is drawn
from the latter in a bootstrap-like manner.  Using these conventions,
\bl
\begin{align*} 
\E_G \hsp d(G,\empk) 
& = \E_G \hsp s(G,\empk) - \E_G \hsp s(\empk,\empk) \\ 
& = \E_G \left[ \frac{1}{k} \sum_{i=1}^k s(G,Y_i) \right] - \E_G \hsp s(\empk,\empk) \\ 
& = s(G,G) - \E_G \hsp s(\empk,\empk) \\ 
& \leq s(F,G) - \E_G \hsp s(\empk,\empk) \\ 
& = \E_G \hsp d(F,\empk), 
\end{align*} 
\el
which is the desired expectation inequality.
\end{proof}

A score divergence in fact satisfies the defining inequality
(\ref{eq:k-proper}) whenever the observations $y_i, \ldots, y_n$
are identically distributed, with or without an assumption of
independence.  In practice, the score divergence
(\ref{eq:score.d}) and the raw score (\ref{eq:score}) yield
the same rankings, though the divergence might be preferred, as it is
nonnegative and anchored at an ideal value of zero.

A usefully general construction proceeds as follows.  Let $h: \Omega \times
\Omega \to [0,\infty)$ be a negative definite kernel,
i.e., $h(x,y) = h(y,x)$ for all $x,y \in \Omega$, and for all positive
integers $n$, all $a_1, \ldots, a_n \in \real$ that sum to zero, and
all $x_1, \ldots, x_n \in \Omega$ it is true that 
\bl
\[
\textstyle \sum_{i=1}^n \sum_{j=1}^n a_i \hsp a_j \hsp h(x_i,x_j) \leq 0.
\]
\el
Subject to natural moment conditions a proper scoring rule can be defined as 
\bl
\begin{equation}  \label{eq:h}
s(F,y) = \E_F h(x,y) - \frac{1}{2} \, \E_F h(x,x'),  
\end{equation}
\el
where $x$ and $x'$ are independent random variables with distribution
$F$ \cite{GneitingRaftery2007}.  In the subsequent sections, we will identify 
various popular divergence functions as score divergences associated with proper 
scoring rules of this form.

Clearly, if the divergence $d$ is $k$-proper for all positive integers
$k$, then it is asymptotically proper.  However, asymptotic propriety
arises under rather weak conditions.

\bigskip 
\noindent{\bf Theorem 3.}  
{\em If for each\/ $G \in \cF$ there exists a constant\/ $c_G$ such
that\ $d(G,\empk) \leq c_G$ for all positive integers\/ $k$ and\/ $d(G,\empk) \to
0$ almost surely, then\/ $d$ is asymptotically proper.}

\begin{proof} 
By the Lebesgue Dominated Convergence Theorem, 
\bl
\[
\textstyle
\lim_{k \to \infty} \E_G \, d(G,\empk) = \E_G \lim_{k \to \infty} d(G,\empk) = 0.  
\]
\el
In view of the divergence $d$ being nonnegative, the defining inequality
(\ref{eq:asymp-proper}) holds for all probability distributions 
$F, G \in \cF$.
\end{proof}

In applications, the number $k$ is often naturally limited by the
number of time points or spatial locations at a fixed scale within the
study domain.  It can then be difficult to assess
whether the value of $k$ at hand is sufficiently large for asymptotic
propriety to be relevant.  Therefore, it is advisable to rank
competing models by using a divergence that is $k$-proper, rather than
just asymptotically proper.  In particular, we advocate the use of
score divergences, as they are guaranteed to be $k$-proper for all
integers $k \geq 1$.


\section{Continuous Outcomes}  \label{sec:continuous} 

We now discuss the propriety of commonly used divergence functions in
the case of continuous outcomes or observations.  In this setting we
assume that $\Omega \subseteq \real^m$, and we let $\cF_p$ denote the
convex class of the Borel probability measures on $\Omega$ with finite
absolute moments of order $p > 0$.

\subsection{Integrated Quadratic Distance}  \label{sec:IQ} 

We initially consider real-valued outcomes and identify a
probability distribution $F$ on $\Omega = \real$ with its cumulative
distribution function (CDF).  The {\em integrated quadratic distance}
is then defined as the integral 
\bl
\begin{equation}  \label{eq:IQ}
d_{\rm IQ}(F,G) = \int_{-\infty}^\infty \left( F(t) - G(t) \right)^2 \dd t, 
\end{equation}
\el
over the squared difference between the corresponding CDFs, when
evaluated at all thresholds.  

If we restrict attention to the class $\cF_1$ of the probability
measure with finite first moment, well known results
\cite{SzekelyRizzo2005, GneitingRaftery2007} allow us to identify the
integrated quadratic distance as the score divergence associated with a kernel
score of the form (\ref{eq:h}), where $h(x,y) = |x-y|$, namely
the {\em continuous ranked probability score}\/
\cite{MathesonWinkler1976, GneitingRaftery2007}, which is defined by
\bl
\begin{equation} 
s(F,y) 
= \int_{-\infty}^\infty \! \left( F(x) - \mathbbm{1} \{y \leq x \} \right)^2 \dd x 
=  \E_F |x - y| - \frac{1}{2} \, \E_F |x - x'|,  \label{eq:c} 
\end{equation}
\el
where $\mathbbm{1} \{y \leq x \}$ denotes an indicator function.  This
proves the $k$-propriety of the integrated quadratic distance for all
positive integers $k$.  Indeed, the score satisfies all the properties
listed in the introduction, with property (i) being immediate from equation
(\ref{eq:IQ}) and property (iii) being implied by the kernel score
representation (\ref{eq:c}).  It also follows 
that we can write
\bl
\[
d_{\rm IQ}(F,G) = \E_{F,G} |x - y| - \frac{1}{2} \, \Big[ \E_F |x - x'|
+ \E_G |y - y'| \Big],
\]
\el
which decomposes $d_{\rm IQ}$  into a term
describing the variability between $F$ and $G$, and
terms relating to the variability within each of $F$ and $G$.

Salazar et al.$\;$\cite{Salazar&2011} apply the continuous ranked
probability score to rank climate model predictions based on the
corresponding mean score of the form (\ref{eq:score}), 
where $F$ is the simulated distribution.  As noted, this yields the
same ranking as applying the integrated quadratic distance
(\ref{eq:IQ}) to the empirical measure of the observations.

A possible generalization is to the class of weighted integrated
quadratic distances of the form
\bl
\[
d_{\rm WIQ}(F,G) = \int_{-\infty}^{\infty} \left( F(t) - G(t) \right)^2 w(t) \: \dd t,
\]
\el
where $w$ is a nonnegative weigth function with
$\int_{-\infty}^{\infty} w(t) \: \dd t < \infty$.  When $F$ and $G$
belong to the class $\cF_1$ it follows from results in 
\cite{MathesonWinkler1976} that $d_{\rm WIQ}$ is a
score divergence and thus it is $k$-proper for all positive integers
$k$.
We may extend further  to higher dimensions, by defining
\bl
\[
d(F,G) = \int_\Omega \left( F(u) - G(u) \right)^2 w(u) \: \dd u,
\]
\el
where $\Omega \subseteq \real^m$, $w$ is a nonnegative weigth function
with $\int_\Omega w(u) \: \dd u < \infty$, and where again we think of
$F$ and $G$ as CDFs.

\subsection{Mean Value Divergence and the Optimal Fingerprint Method}  \label{sec:Mahalanobis} 

In the optimal fingerprint method \cite{Hasselmann1993,
Hegerl&1996, Barnett&2005, Zhang&2007} a model generated
climate change signal is said to be detected if its amplitude in the
observations is large compared to the amplitude of unforced model
output signal or uncontaminated data.  The amplitude is measured by
the square of the Mahalanobis distance \cite{Mahalanobis1936} using
the covariance matrix of the natural climate variability.  In our
setting, we consider the divergence between probability measures $F$
and $G$ on $\real^m$ that is defined by the square,
\bl
\[
d(F,G) = ( \hsp \mu_F - \mu_{G})' \, \Sigma^{-1} ( \hsp \mu_F - \mu_{G}), 
\]
\el
of the Mahalanobis distance, where the positive definite matrix
$\Sigma \in \real^{m \times m}$ may depend on $G$ but not on $F$.
This distance depends on the predictive distribution
$F$ only through the mean vector $\mu_F \in \real^m$, and it is readily
seen that
\bl
\[
\E_G \hsp d(F,\empk) - \E_G \hsp d(G,\empk) 
= d(F,G) 
\geq 0,
\]
\el
whence $d$ is $k$-proper for all positive integers $k$.  In
particular, if we take $\Sigma$ to be the identity matrix, we obtain
the {\em mean value divergence},
\bl
\begin{equation}  \label{eq:MV}
d_{\rm MV}(F,G) = ( \hsp \mu_F - \mu_G )' \hsp ( \hsp \mu_F - \mu_G).   
\end{equation}
\el
The variant defined by $d(F,G) = (\mu_F - \mu_{G})' \, \Sigma_{F}^{-1}
(\mu_F - \mu_{G})$, where $\Sigma_F \in \real^{m \times m}$ is the
covariance matrix of $F$, fails to be proper, as we can diminish
the expected divergence by artificially inflating $\Sigma_F$.

\subsection{Dawid-Sebastiani Divergence}  \label{sec:DS} 

Dawid and Sebastiani \cite{DawidSebastiani1999} study score
divergences between probability measures $F$ and $G$ on $\real^m$ that
depend on the mean vector, $\mu_F \in \real^m$, and the covariance
matrix, $\Sigma_F \in \real^{m \times m}$ of the forecast distribution
only.  A prominent example is the {\em Dawid-Sebastiani divergence},
\bl
\begin{equation}  \label{eq:DS}
d_{\rm DS}(F,G) 
= \, \textup{tr}(\Sigma_{F}^{-1}\Sigma_{G}) - \log \det(\Sigma_F^{-1}\Sigma_{G})  
+ (\mu_F - \mu_{G})' \, \Sigma_{F}^{-1} (\mu_F - \mu_{G}) - m, 
\end{equation}
\el
which arises as the score divergence associated with the proper scoring
rule \cite{DawidSebastiani1999, GneitingRaftery2007}
\bl
\[
s(F,y) = \log \det \Sigma_F + (\mu_F - y)' \, \Sigma_{F}^{-1} (\mu_F - y).   
\]
\el
The expression on the right-hand side corresponds to  the non-normalized
log-likelihood function for a multivariate normal density, whence
$d_{\rm DS}(F,G)$ is equivalent to the Kullback-Leibler divergence
between multivariate normal distributions with mean vectors and
covariance matrices equal to those of $F$ and $G$.

\subsection{Area Validation Metric and Wasserstein Distance}  \label{sec:AVM} 

We now return to real-valued outcomes and identify a probability
distribution $F$ on $\Omega = \real$ with its CDF.  In this setting, 
Ferson et al.$\,$\cite{Ferson&2008} 
propose the {\em area validation metric},
\bl
\[
d_{\rm AV}(F,G) = \int_{-\infty}^\infty |F(t)-G(t)| \, \dd t, 
\]
\el
as a divergence function.  This resembles the integrated quadratic distance 
(\ref{eq:IQ}) and satisfies the desirable properties (i), (ii) and (iii) 
stated in the introduction.  

Furthermore, by Theorem 3 and standard results in the theory of
empirical processes \cite[p.~66]{ShorackWellner1986}, the area validation
metric is asymptotically proper as a divergence measure for
probability distributions with finite first moment.  Nevertheless,
property (iv) is violated, as $d_{\rm AV}$ generally fails to be $k$-proper.  
For example, let $G$ be the uniform distribution on $[0,1]$, and let $F_k$ be discrete 
with probability mass $1/k$ in $x = i/(k+1)$ for $i = 1, \ldots, k$.  Then
\bl
\[ 
\textstyle
\frac{1}{4} = \E_G \hsp d_{\rm AV}(F_1,\emps) 
            < \E_G \hsp d_{\rm AV}(G,\emps) = \frac{1}{3}, 
\]
\el
and $\E_G \hsp d_{\rm AV}(F_k,\empk) < \E_G \hsp d_{\rm AV}(G,\empk)$
for $k \leq 25$ in Monte Carlo experiments, thereby suggesting that
$d_{\rm AV}$ encourages underdispersed model simulations.

The area validation metric can be identified with a special case of the 
{\em transportation distance}\/ or {\em Wasserstein distance}\/ 
\cite{BickelFreedman1981, Deza&2013} of order $p \geq 1$, 
namely 
\bl
\[
d_p(F,G) 
= \left( \! \rule{0mm}{5mm} \inf \, \E_{F,G} \hsp |x-y|^p \right)^{\! 1/p}
= \left( \, \int_0^1 | F^{-1}(u) - G^{-1}(u)|^p \: \dd u \right)^{\! 1/p}
\]
\el
where $F^{-1}(u) = \inf \{ t : F(t) \geq u \}$ and the infimum is
taken over all jointly distributed random variables $x$ and $y$ with
marginal distributions $F$ and $G$, respectively.   For
$p=1$, it holds that $d_1(F,G) = d_{\rm AV}(F,G)$,  and
for $p=2$ we may write $d_2^2(F,G) = (\mu_F - \mu_G)^2 +
d_2^2(F_0,G_0)$, where $F_0$ and $G_0$ are shifted versions of $F$ and
$G$ with mean zero \cite{BickelFreedman1981}.  While the Wasserstein
distance $d_p$ is asymptotically proper as a divergence measure within
the class $\cF_p$, it generally fails to be $k$-proper for finite $k$.

\subsection{Kolmogorov-Smirnov Distance}  \label{sec:KS} 

To continue the consideration of real-valued outcomes, the {\em Kolmogorov-Smirnov 
distance},
\bl
\[
\textstyle
d_{\rm KS}(F,G) = \sup_{t \in \real} |F(t) - G(t)|, 
\]
\el
where $F$ and $G$ are interpreted as CDFs, is commonly used as a
divergence function.  By the Glivenko-Cantelli
Theorem and our Theorem 3, the Kolmogorov-Smirnov distance is
asymptotically proper.  However, it is generally not $k$-proper. 
For example, if $G$ is uniform on $[0,1]$ and
$F_k$ is discrete with mass $1/(k+1)$ in $i/k$ for $i
= 0, \ldots, k$, then
\bl
\[ 
\textstyle
\frac{1}{2} = \E_G \hsp d_{\rm KS}(F_1,\emps) 
            < \E_G \hsp d_{\rm KS}(G,\emps) = \frac{3}{4}, 
\]
\el
and $\E_G \hsp d_{\rm KS}(F_k,\empk) < \E_G \hsp d_{\rm KS}(G,\empk)$
for $k \leq 5$ in Monte Carlo experiments.  We thus have
reservations about the use of the Kolmogorov-Smirnov distance for
ranking competing models; however, we hasten to add that any concerns 
about the use of improper divergence functions are task dependent and 
may not apply to testing problems.

\section{Categorical Outcomes}  \label{sec:examples} 

We now consider categorical outcomes or observations.  Without loss of
generality, we may assume that the sample space is the finite set
$\Omega = \{ 1, \ldots, c \}$.  A probability distribution $F$ on
$\Omega$ can then be identified with a probability vector $(f_1,
\ldots, f_c)$.  Similarly, the empirical measure $\empk$ can be
identified with a vector of the form $(\widehat{g}_{k1}, \ldots,
\widehat{g}_{kc})'$, where $\widehat{g}_{kj} = \#\{ y_i = j : i = 1,
\ldots, k \} / k$ for $j = 1, \ldots, c$.  This is
frequently the setting in climate model studies, where simulations
and observations typically are continuous values, but may be binned or
categorized prior to further analysis.

\subsection{Kullback-Leibler Divergence}  \label{sec:KL} 

The {\em Kullback-Leibler diver\-gence}\/ is a commonly used
asymmetric measure of the difference between two probability vectors,
namely
\bl
\begin{equation}  \label{eq:KL}
d_{\rm KL}(F,G) = \sum_{i=1}^c f_i \log \frac{f_i}{g_i}.  
\end{equation}
\el
The Kullback-Leibler divergence is the score divergence associated with 
the proper logarithmic score \cite{GneitingRaftery2007, BrockerSmith2007} 
and so it is $k$-proper for all positive integers $k$ by Theorem 2.

\subsection{Brier Divergence}  \label{sec:Brier} 

The {\em quadratic}\/ or {\em Brier divergence}, 
\bl
\begin{equation}  \label{eq:quadratic}
d_{\rm B}(F,G) = \sum_{i=1}^c \left( f_i - g_i \right)^2  
\end{equation}
\el
is the score divergence associated with the Brier score
\cite{Brier1950, GneitingRaftery2007}, and thus it is $k$-proper for
all positive integers $k$.

\subsection{Hellinger Distance}  \label{sec:Hellinger} 

Palmer \cite{Palmer2012} suggests the use of the {\em Hellinger distance},
\bl
\[
d_{\rm H}(F,G) = \left( 
  \frac{1}{2} \sum_{i=1}^c \left( \sqrt{f_i} - \sqrt{g_i} \, \right)^{\! 2} \right)^{\! 1/2}, 
\]
\el
to compare probability distributions derived from climate model output
to the corresponding data distributions.  While the Hellinger distance
is asymptotically proper by Theorem 3, it generally fails to be
$k$-proper.  When $c = 2$ and $F$ and $G$ are represented by the probability 
vectors $(f_1,1-f_1)$ and $(g_1,1-g_1)$, respectively, we find that 
\bl
\[
\E_G \hsp d_{\rm H}(F,\emps) = g_1 \sqrt{1-\sqrt{f_1}} + (1-g_1) \sqrt{1 - \sqrt{1 - f_1}}, 
\] 
\el
which generally does not attain a minimum when the forecast probability $f_1$ equals $g_1$.  
For instance, if $f_1 = 0.10$ and $g_1 = 0.25$, Monte Carlo experiments show that 
\bl
\[ 
\E_G \hsp d_{\rm H}(F,\empk) < \E_G \hsp d_{\rm H}(G,\empk)
\] 
\el
for $k = 1, 2, 5, 6$ and $10$.  

\section{Case study: Temperature Extremes Over Europe}  \label{sec:casestudy}

Anthropogenic climate change has awaken strong scientific, societal
and policy interest in predicting future climate evolution.  The
simulation of climate elements is usually carried out by
coupled atmosphere-ocean circulation models.  The output from the
models is then used to provide insight into future climate states.  In
this illustration, we focus on annual maxima of 2m (surface)
temperature.  We evaluate 15 different climate models by comparing
simulations of past climate to corresponding re-analysis data, where
historical weather observation data are used to reconstruct realized
states of the atmosphere on a global grid, thereby facilitating direct
comparison to climate model output.

\begin{figure}
\centering
\subfigure[IQ distance from ERA-40\label{ERA40_dfcrps}]{
\includegraphics[width=0.4\textwidth]{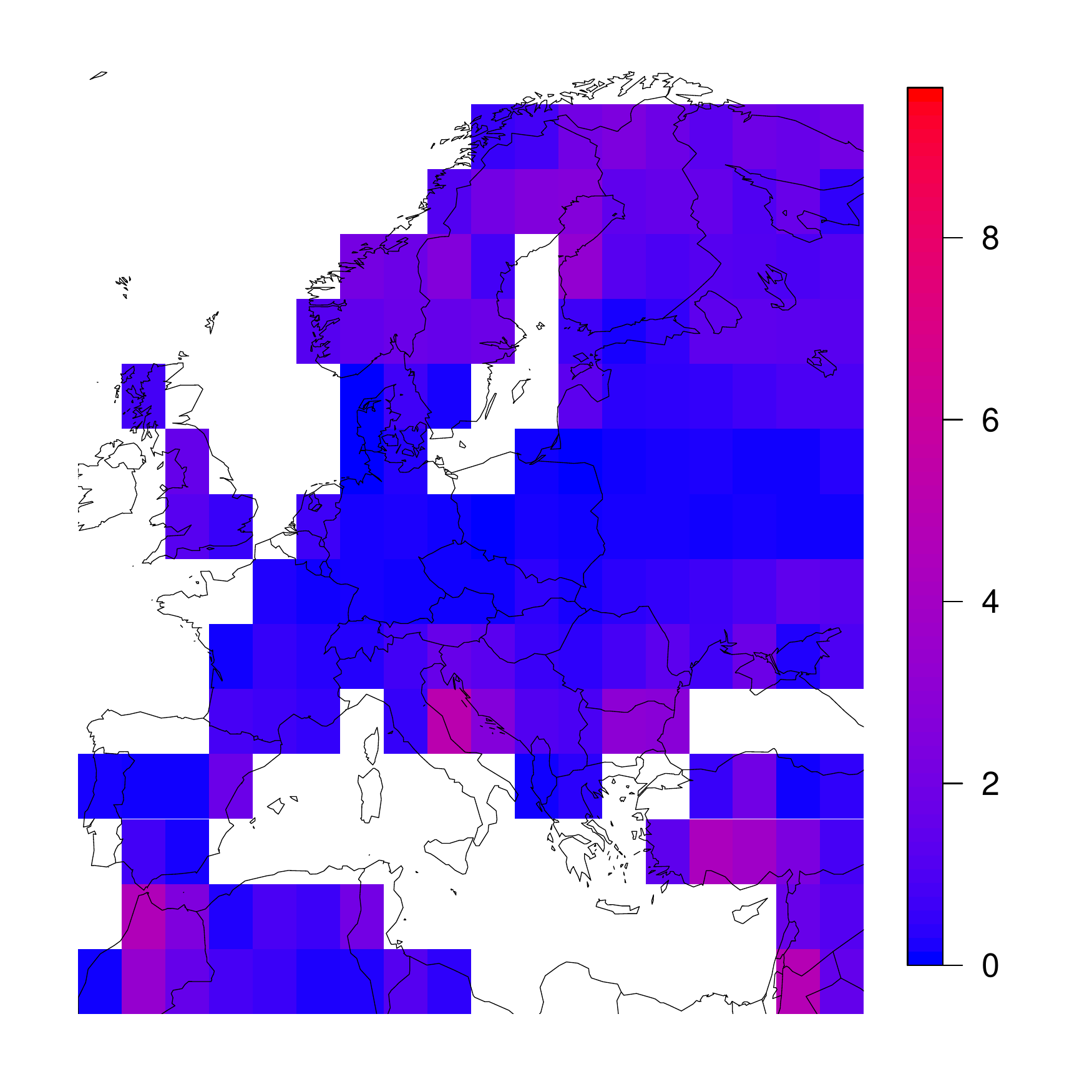}}
\quad
\subfigure[IQ distance from NCEP-1\label{NCEP_dfcrps}]{
\includegraphics[width=0.4\textwidth]{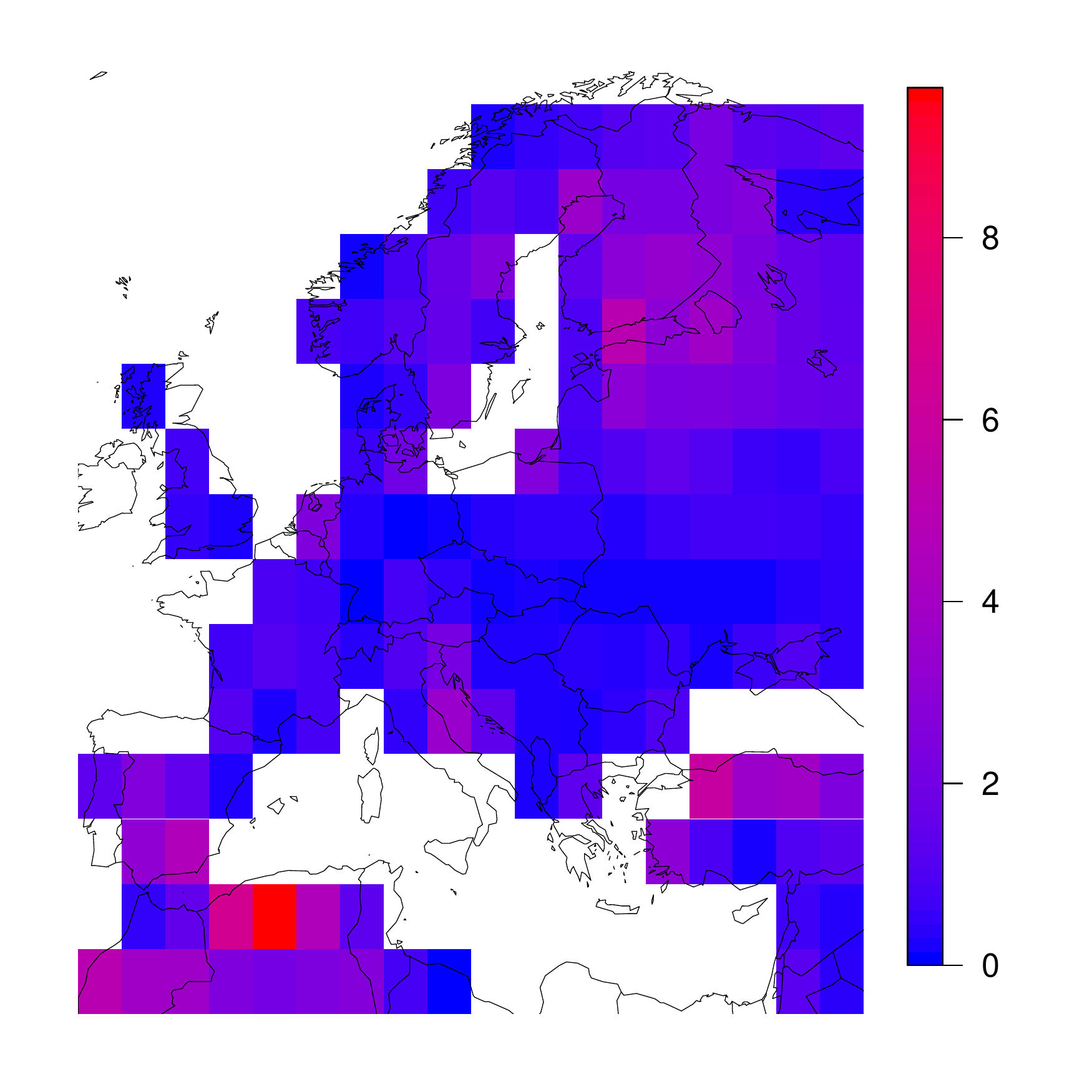}}
\subfigure[MV divergence from ERA-40 \label{ERA40_smvd}]{
\includegraphics[width=0.4\textwidth]{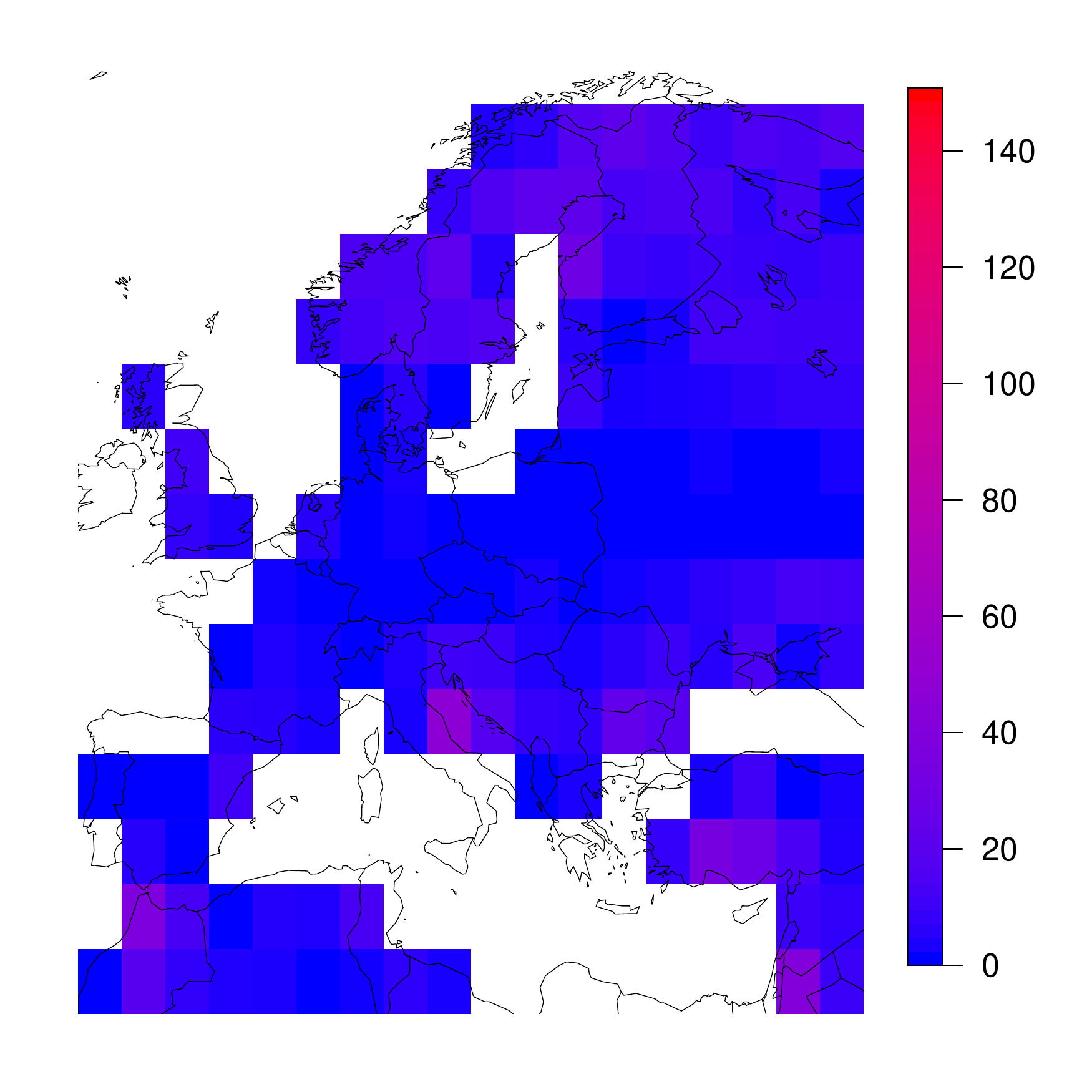}}
\quad
\subfigure[MV divergence from NCEP-1 \label{NCEP_smvd}]{
\includegraphics[width=0.4\textwidth]{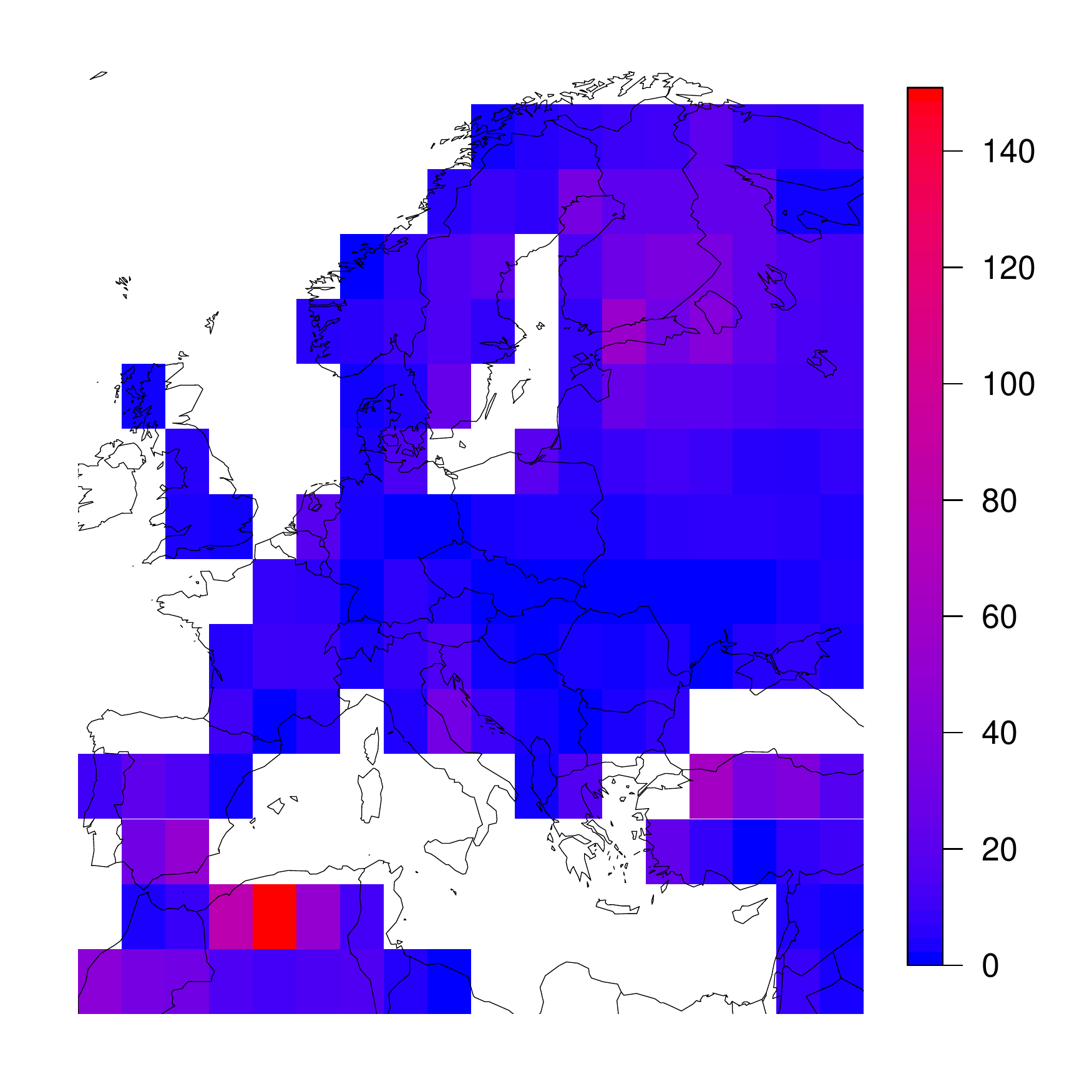}}
\caption{Grid box specific divergences of yearly maximum temperature
  simulations for the period 1961--1990 by the ECHAM5/MPI-OM model in
  terms of the integrated quadratic (IQ) distance, in the unit of degrees Celsius, and 
  the mean value (MV) divergence from the re-analysis data ERA-40 and NCEP-1.}
\label{fig:divmes}
\end{figure}

Our study region consists of large parts of Europe (8W--40E, 32N--74N)
as illustrated in Figure \ref{fig:divmes}.  The climate model
projections are obtained from the Coupled Model Intercomparison
Project phase 3 (CMIP3) multi-model dataset \cite{Meehl&2007}.  The
various models operate on distinct global grids.  To allow for a grid
based comparison, we transform to a common global grid of size $144 \times
73$, as used for the re-analysis data.  At this resolution, our study
region contains a total of 154 grid boxes.  As observational data,
we work with the ERA-40 re-analysis \cite{Uppala&2005} by the European
Centre for Medium-Range Weather Forecasts and the NCEP-1 re-analysis
\cite{Kalnay&1996} by the U.S. National Centers for Environmental
Prediction.  In our comparison, we only use a single simulation from
each of the 15 climate models; if multiple simulations are available,
one instance is selected at random.  Both the climate models and the
re-analyses provide daily outputs for 2m maximum temperature.  
At each grid box, the climate model simulation then is represented by 
the location-specific empirical distribution function of the resulting
annual maxima over the study period from 1961 to 1990.

\begin{figure} 
\centering
\includegraphics[width=\textwidth]{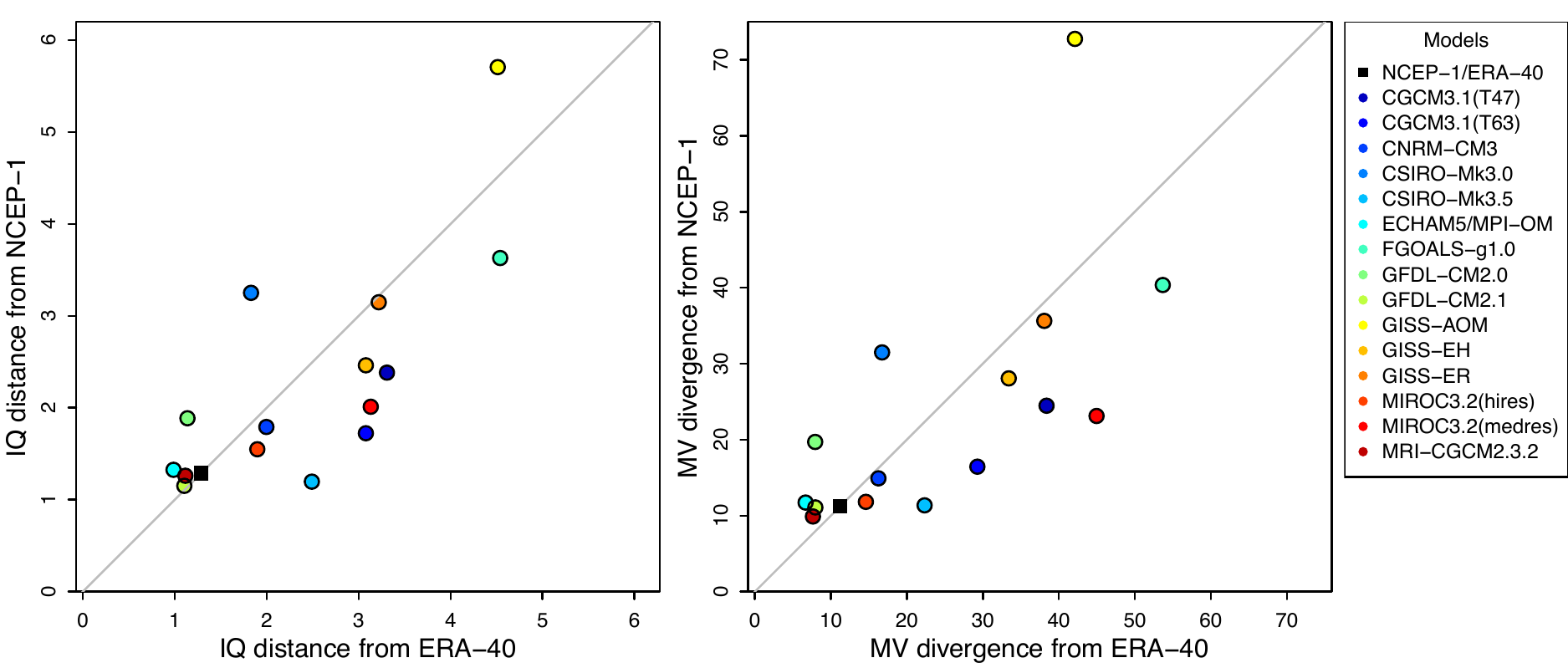}
\caption{ Evaluation of yearly maximum temperature
  simulations for the period 1961--1990 by the 15
  CMIP3 climate models in terms of the integrated quadratic (IQ)
  distance, in the unit of degrees Celsius, and  in
  terms of the mean value (MV) divergence, from the re-analysis data
  ERA-40 (x-axis) and NCEP-1 (y-axis).
  The values correspond to the average divergence over the 154 grid
  cells in Figure \ref{fig:divmes} and have been jiggered
    to improve clarity. }
\label{fig:allyear} 
\end{figure}

Figure \ref{fig:allyear} shows the integrated quadratic distance
(\ref{eq:IQ}) in the unit of degrees Celsius and the mean value
divergence (\ref{eq:MV}) of the simulations from the two sets of
re-analysis data, with each value being an average over the 154 grid
boxes.  The divergences result in similar rankings of
the climate models relative to the re-analysis data sets.
For three models, the divergences between the model and the
re-analysis data are at the level of the internal variability between
the ERA-40 and the NCEP-1 re-analyses, which is the best performance
one might reasonably hope for.  These are the ECHAM5/MPI-OM model from
the Max Planck Institute for Meteorology in Germany, the GFDL-CM2.1
model developed by the Geophysical Fluid Dynamics Laboratory in the
United States and the MRI-CGCM2.3.2 model from the Meteorological
Research Institute of Japan.  The first two models have a slightly
higher spatial resolution than the re-analyses while the MRI-CGCM2.3.2
model operates on a grid of size $128 \times 64$.

In Figure \ref{fig:divmes}, we can identify the
location-specific divergences for the ECHAM5/MPI-OM model under both
sets of re-analysis data.  Any differences between the two re-analyses
seem to be more pronounced in coastal regions, in particular around the
Mediterranean Sea and on the Eastern shore of the Baltic Sea.  As the
two types of divergences operate on different scales, the results for
the integrated quadratic distance cannot be compared directly to the
mean value divergence.  However, the results seem to indicate that
there is more local variability under the integrated quadratic
distance, as is to be expected, given that it addresses all
distributional features, as opposed to the mean value divergence.

\section{Discussion}  \label{sec:discussion}

We have introduced the notion of propriety for divergences, i.e.,
distance measures between probability distributions.  In a nutshell,
if the divergence is proper, and we believe that the observations form
a sample from the probability distribution $G$, then an optimal,
expectation minimizing predictive distribution is $G$.  This property
is important in settings such as that of Figure 2, where we compared
climate models based on the average divergence between the probability
distributions of model output and the corresponding empirical
distributions of temperature extremes.  Empirical means
correspond to expectations and therefore, for a proper divergence
function, the average divergence criterion favors skillful climate
models in decision theoretically coherent ways.  In this context, the
question for meaningful or perhaps even optimal ways of averaging and
aggregating deserves further study.
 
Score divergences tha t derive from proper scoring rules are
$k$-proper for all positive integers $k$, with the integrated
quadratic distance (\ref{eq:IQ}), the Dawid-Sebastiani divergence
(\ref{eq:DS}) and the Kullback-Leibler divergence (\ref{eq:KL}) being
particularly attractive choices, while other commonly used distances
fail to be proper.  In the case of real-valued variables, the
integrated quadratic distance is the only divergence available that
satisfies the desirable properties listed in the introduction, and
thus we endorse and encourage its use.

The evaluation of model simulations continues to provide challenges.
In this paper we have considered the case of individual
climate model runs, which we have ranked based on their grid averaged
divergences from re-analysis data sets.  Fricker et al.~\cite{Fricker&2013}
consider further simulation formats, such as ensemble hindcasts, and
in the case $m = 1$ of a single model run, their pooled ensemble CRPS
\cite[p.~252]{Fricker&2013} equals the integrated quadratic distance.
An important caveat is that if we evaluate model simulations for a
univariate quantity, such as an annual temperature maximum, by using a
divergence function, the assessment is restricted to univariate
aspects.  If multivariate aspects, such as the behavior of temperature
extremes at several sites simultaneously, are of interest, divergence
functions for probability distributions in higher-dimensional spaces
are to be considered.

\section*{Acknowledgments} 

The work of Nadine Gissibl and Tilmann Gneiting has been supported by
the European Union Seventh Framework Programme under grant 
agreement no. 290976, Thordis L. Thorarinsdottir 
acknowledges the support of sfi$^2$, Statistics for Innovation in Oslo.
We are grateful to Sebastian
Mieruch, Jana Sillmann, Jon Wellner and  Johanna
Ziegel for discussions and/or assistance with the
data handling, and thank an associate editor and two
reviewers for constructive comments.  We acknowledge the
modeling groups, the Program for Climate Model Diagnosis and
Intercomparison (PCMDI) and the World Climate Research Programme's
(WCRP's) Working Group on Coupled Modelling (WGCM) for making the WCRP
CMIP3 multi-model dataset available.  Support of this dataset is
provided by the Office of Science, U.S. Department of Energy.

\end{document}